\documentclass[letterpaper,prl,twocolumn,showpacs]{revtex4}

\usepackage{amsmath,amssymb}
\usepackage{epsfig}

\newcommand {\ra}{\rightarrow}

\begin{document}

\title{Soliton on Unstable Condensate}

\date{\today}

\author{V.\,E.~Zakharov$^{1,2,3}$} \email{zakharov@math.arizona.edu}
\author{A.\,A.~Gelash$^{2}$} \email{agelash@gmail.com}
\affiliation{$^{1}$Department of Mathematics, University of Arizona, Tucson, AZ, 857201, USA}
\affiliation{$^{2}$Novosibirsk State University, Pirogova, 2, Novosibirsk, 630090, Russia}
\affiliation{$^{3}$Physical Institute of RAS, Leninskiy prospekt, 53, Moscow, 119991, Russia}


\begin{abstract}
We construct new exact solutions of the focusing Nonlinear Schr\"{o}dinger equation (NLSE). This is a soliton propagating on an unstable condensate. The Kuznetsov and Akhmediev solitons as well as the Peregrine instanton are particular cases of this new solution. We discuss applications of this new solution to the description of freak (rogue) waves in the ocean and in optical fibers.
\end{abstract}

\maketitle {\it --Introduction --} This research was motivated by intention to develop an analytic theory of freak (or rogue) waves in ocean and optic fibers. In the recent time the simplest and most universal model for description of these waves is the focusing NLSE (Nonlinear Schr\"{o}dinger Equation). In application to the theory of ocean waves this equation is used since 1968 ~\cite{Zakharov1968}. In nonlinear optics it was known even earlier ~\cite{Townes1964}.

The focusing NLSE is the model of the first approximation only.
For the surface of fluid this model describes the essentially weakly nonlinear quasimonochromatic wave trains with average steepness not more than $5\cdot10^{-2}$ only ~\cite{Dyachenko2008}. In nonlinear optics its application is also limited to the case of waves of small amplitudes (see, for instance ~\cite{Kivshar-Agraval2003}). Nowadays a lot of models specializing the NLSE  are developed. For the surface waves these are Dysthe equations ~\cite{Dysthe-Trulsen1999,Zakharov-Dyachenko2010}, for the waves in optic fibers are equations that include the third time derivatives and more complex forms of nonlinearity (see for instance ~\cite{Zakharov-Kuznetsov1998,Balakin2007}). Also, freak waves in the ocean were studied by numerical modeling of exact Euler equations for potential flow with free boundary ~\cite{Zakharov-Dyachenko-Prokofiev2006,Chalikov-Sheinin2005}. The behavior of freak waves studied by NLSE and by more sophisticated models shows considerable quantitative difference. Nevertheless, advanced improvement of NLSE does not lead to any qualitatively new effects. That means that a careful and detailed study of NLSE solutions is still very important problem.

It has been known since 1971 that the focusing NLSE is a very special system, integrable by the Inverse Scattering Method ~\cite{Zakharov-Shabat1972}. Since that time hundreds of articles have been devoted to this subject (see for instance the monographs ~\cite{Faddeev-Takhtajan2007,Sulem-Sulem1999,Kharif-Pelinovsky2009}. In spite of great attention to this field, solutions of focusing NLSE in the presence of an unstable condensate have not been studied thoroughly enough. However, this case is the most interesting for developing a theory of freak waves, because there is now universal agreement that extreme waves appear as a result of modulation instability of quasimonochromatic nonlinear waves. A nonlinear theory describing the development of unstable condensate in the frame of 1-D focusing NLSE has not yet been formulated. Moreover, even ordinary solitonic solutions propagating over unstable condensate have not yet been found.

Some special cases are already known. In the presence of condensate the maximally analytic wave function of auxiliary linear Zakharov-Shabat operator has, in the right half-plane of the complex variable $\lambda=i\eta$ (where $\eta$ is the spectral parameter), an additional cut on the real axis up to the point $\lambda=A$ (here $A$ is the condensate amplitude). The soliton adds a simple pole to the wave function singularity. Only the cases when the pole is on the real axis have been studied. In 1978 Kuznetsov ~\cite{Kuznetsov1977} found an important solitonic solution with the pole outside of the cut $Re \lambda >A$. This solution is a standing soliton with a changing amplitude. Later on this solution was rediscovered by other authors ~\cite{Kawata1978,Ma1979}.
In 1983 Peregrine~\cite{Peregrine1983} found a solution with the pole exactly at the branch point. This is a remarkable instanton homoclinic solution - the localized events appear from the condensate and return back to it. Now there are already known multi-instanton solutions ~\cite{Matveev2010,Matveev2011}. The importance of these solutions for the development of the theory of freak waves is stressed in the article ~\cite{Shrira2010}. Then, the solution of Akhmediev with co-authors is widely known ~\cite{Akhmediev1986}. This solution corresponds to the pole posed inside the cut and is periodic in space but local in time. Unlike the Peregrine instanton, this solution change the condensate phase.

In this article we describe a general solitonic solution with the pole at an arbitrary point in the complex upper half-plane of the spectral parameter. This solution includes all other known cases and changes the condensate phase. It is a moving localized soliton with periodic nonmonochromatic intrinstic structure. When the pole is far away from the cut, but close to real axis, the solution is similar to Kuznetsov's soliton. However, when approaching the cut (but still not on the real axis) the velocity of soliton propagation and the size of soliton tend to infinity, so the Akhmediev limit happens to be very singular.

For constructing this solution we use the advanced mathematical technique of the "local $\overline{\partial}$ - problem". This method is a modernization of the "dressing method", described in ~\cite{Zakharov-Shabat1979,Zakharov-Manakov-Novikov-Pitaevskii1984}.

{\it --NLSE via local matrix $\overline{\partial}$ problem --}
We study solutions of the following NLSE
\begin{equation}
i\varphi_{t}-\frac{1}{2}\varphi_{xx}-(|\varphi|^{2}-A^{2})\varphi=0 \label{NLSE}
\end{equation}
with boundary conditions $|\varphi|^2\rightarrow A^{2}$ at $x\rightarrow\pm\infty$. Here $A=\overline{A}$ is a real constant. Equation (\ref{NLSE}) is the compatibility condition for the following overdetermined linear system for a matrix function $\Psi$:
\begin{equation}
\frac{\partial\Psi}{\partial x}=\widehat{U}\Psi ,\;\;
i\frac{\partial\Psi}{\partial t}=(\lambda\widehat{U}+\widehat{W})\Psi \label{lax system}
\end{equation}
Here
\begin{eqnarray}
\widehat{U}=I\lambda+u
\nonumber\\
I=\begin{pmatrix}1 & 0\\ 0 & -1\end{pmatrix}
,\;\;
u=\begin{pmatrix}0 & \varphi\\ -\overline{\varphi} & 0\end{pmatrix}
\nonumber\\
\widehat{W}=\frac{1}{2}\begin{pmatrix}|\varphi|^{2}-|A|^{2} & \varphi_{x}\\ \overline{\varphi}_{x} & -|\varphi|^{2}+|A|^{2}\end{pmatrix}
\label{U and W def.}
\end{eqnarray}
If $\varphi=0$, system (\ref{lax system}) has the following solution:
\begin{equation}
\Psi_{0}=\frac{1}{\sqrt{1-q^{2}}}\begin{pmatrix}e^{\phi} & q\cdot e^{-\phi}\\ q\cdot e^{\phi} & e^{-\phi} \end{pmatrix} \label{condensat lax solution}
\end{equation}
Here
\begin{eqnarray}
\phi=kx+\Omega t ,\;\; k^{2}=\lambda^{2}-A^{2}
\nonumber\\
\Omega=-i\lambda k ,\;\; q=-\frac{A}{\lambda+k}
\nonumber
\end{eqnarray}
In what follows we assume that $k \ra \lambda$ at $|\lambda| \ra \infty$. Then
\begin{equation}
\Psi_{0}^{-1}=\frac{1}{\sqrt{1-q^{2}}}\begin{pmatrix}e^{-\phi} & -q\cdot e^{-\phi}\\ -q\cdot e^{\phi} & e^{\phi} \end{pmatrix} \label{condensat lax solution^-1}
\end{equation}
Notice that
\begin{eqnarray}
\overline{k}(-\overline{\lambda})=-k(\lambda)
,\;\;\;\; \overline{q}(-\overline{\lambda})=-q(\lambda)
,\;\;\;\; \overline{\phi}(-\overline{\lambda})=\phi(\lambda)
\nonumber
\end{eqnarray}
and
\begin{equation}
\Psi^{-1}_{0}(-\overline{\lambda})=\Psi^{+}(\lambda) \label{Psi^-1 and Psi^+ eq.}
\end{equation}
We consider the $\overline{\partial}$-problem on the complex $\lambda$-plane. We are looking for a second order matrix  $\chi(\lambda,\overline{\lambda},x,t)$ obeying the equation
\begin{equation}
\frac{\partial\chi}{\partial\overline{\lambda}}=\chi\cdot f(\lambda,\overline{\lambda},x,t) \label{delta with line problem}
\end{equation}
and normalized by condition $\chi\rightarrow1$ at $|\lambda|\rightarrow\infty$.
Here
\begin{equation}
f=\psi_{0} f_{0}(\lambda,\overline{\lambda})\psi_{0}^{-1}
\nonumber
\end{equation}
\begin{equation}
f_{0}(\lambda,\overline{\lambda})=f^{+}_{0}(\lambda,\overline{\lambda}) \label{f^-1 and f^+ eq.}
\end{equation}
By virtue of (\ref{Psi^-1 and Psi^+ eq.}) function $f$ satisfies condition $f(\lambda,\overline{\lambda})= f^{+}(\lambda,\overline{\lambda})$. This proves that
\begin{equation}
\chi^{-}(\lambda)=\chi^{+}(-\overline{\lambda}) \label{chi^-1 and chi^+ eq.}
\end{equation}
The function $\chi$ has an asymptotic expansion at $\lambda \ra \infty$
\begin{equation}
\chi=1+\frac{R}{\lambda}+\cdots
\end{equation}
By virtue of (\ref{f^-1 and f^+ eq.}) $R^{+}=R$. The function $\chi$ satisfies the following system of equations
\begin{eqnarray}
\frac{\partial \chi}{\partial x}=\widehat{U}\chi-\chi \widehat{U}_{0} \nonumber\\
i\frac{\partial \chi}{\partial t}=(\lambda \widehat{U}+\widehat{W})-\lambda\chi\widehat{U_{0}} \label{chi system}
\end{eqnarray}
Here $\widehat{U}$ and $\widehat{W}$ are given by expressions (\ref{U and W def.}). Because system (\ref{chi system}) is overdeterminated, we have the following expression for $\varphi$:
\begin{equation}
\varphi=A-2R_{(12)}
\end{equation}
For an arbitrary choice of matrix function $f_{0}(\lambda,\overline{\lambda})$ satisfying condition (\ref{f^-1 and f^+ eq.}), function $\varphi$ is the solution of equation (\ref{NLSE}).

\maketitle {\it --Solitonic solution --}
Let us choose
\begin{equation}
f_{0}(\lambda,\overline{\lambda})=
\begin{pmatrix}0 & \alpha(\lambda,\overline{\lambda})\\ \overline{\alpha}(-\overline{\lambda},-\lambda) & 0\end{pmatrix} \nonumber
\end{equation}
The function $f$ now takes the form:
\begin{equation}
f(\lambda,\overline{\lambda},x,t)=
\alpha e^{2\phi}A+\overline{\alpha}(-\overline{\lambda},-\lambda)Be^{-2\phi}
\nonumber
\end{equation}
The matrices $A,B$ are degenerate
\begin{equation}
A_{\alpha\beta}=a_{\alpha}b_{\beta} ,\;\; B_{\alpha\beta}=c_{\alpha}d_{\beta}
\nonumber
\end{equation}
\begin{eqnarray}
a=(1,q) ,\;\; b=(-q,1)
;\;\;\;\;
c=(q,1) ,\;\; d=(1,-q)
\nonumber
\end{eqnarray}
and
\begin{eqnarray}
(\overrightarrow{a} \cdot \overrightarrow{b})=a_{\alpha}b_{\alpha}=0
,\;\;\;\;\;\;\;\;\;\;\;\;
(\overrightarrow{c} \cdot \overrightarrow{d})=c_{\alpha}d_{\alpha}=0
\nonumber
\end{eqnarray}
We now choose
\begin{equation}
\alpha(\lambda,\overline{\lambda})=C\delta(\lambda-\eta)
\nonumber
\end{equation}
Here $\delta(\lambda-\eta)$ is the two-dimensional $\delta$-function. Later on, the existence of $\delta$-function allows us to work with two values of
 function $\phi$:
\begin{eqnarray}
\phi(\eta)=k(\eta)(x-i\eta t) ,\;\; \phi(-\overline{\eta})=-\overline{\phi}
\nonumber
\end{eqnarray}
We now assume that $q(\eta)=q$, $q(-\overline{\eta})=-\overline{q}$. In a neighborhood of $\lambda=\eta$, $\lambda=-\overline{\eta}$ we expand A and B:
\begin{eqnarray}
A=A_{0}+A_{1}(\lambda-\eta) ,\;\;
B=B_{0}+B_{1}(\lambda+\overline{\eta}) \nonumber\\
A_{0}=\begin{pmatrix}-q &1 \\ -q^{2} &q\end{pmatrix}
,\;\;
B_{0}=\begin{pmatrix}q &-q^{2} \\ 1 &-q\end{pmatrix}
\nonumber\\
A_{1}=-\frac{q}{k}\begin{pmatrix}1 &0 \\ -2q &0\end{pmatrix}
,\;\;
B_{1}=-\frac{\overline{q}}{\overline{k}}\begin{pmatrix}-1 &-2\overline{q} \\ 0 &1\end{pmatrix}
\nonumber
\end{eqnarray}
Note that now $\overrightarrow{c}=\overrightarrow{b^{*}}$ and $\overrightarrow{d}=\overrightarrow{a^{*}}$.
We will find a solution of the $\overline{\partial}$ - problem (\ref{delta with line problem}) in the form of rational functions with two poles:
\begin{equation}
\chi=1+\frac{U}{\lambda-\eta}+\frac{V}{\lambda+\overline{\eta}} \label{chi solution}
\end{equation}
where $U$, $V$ are constant degenerate matrices:
\begin{equation}
U_{\alpha\beta}=u_{\alpha}b_{\beta} ,\;\; V_{\alpha\beta}=v_{\alpha}a^{*}_{\beta}
\nonumber
\end{equation}
To avoid a singularity we require
\begin{equation}
UA_{0}=0 ,\;\; VB_{0}=0
\nonumber
\end{equation}
Substituting (\ref{chi solution}) into (\ref{delta with line problem}) we end up with a simple linear system of equations for $u_{\alpha}$ and $v_{\alpha}$:
\begin{eqnarray}
u_{\alpha}(1+\frac{q}{k}Ce^{2\phi})-\frac{1+|q|^2}{\eta+\overline{\eta}}Ce^{2\phi}v_{\alpha}=
a_{\alpha}Ce^{2\phi}
\nonumber\\
\frac{1+|q|^2}{\eta+\overline{\eta}}Ce^{2\overline{\phi}}u_{\alpha}+(1+\frac{\overline{q}}{\overline{k}}\overline{C}e^{2\overline{\phi}})v_{\alpha}=
\overline{b}_{\alpha}\overline{C}e^{2\overline{\phi}}
\nonumber
\end{eqnarray}
By virtue of (15) the solution of equation (\ref{NLSE}) is given as follows
\begin{equation}
\varphi=A-2(u_{1}+v_{1}\overline{q})
\nonumber
\end{equation}
It is convenient to present final answer in terms of a uniformizing variable $\xi$:
\begin{eqnarray}
\lambda=\frac{A}{2}(\xi+\frac{1}{\xi}) ,\;\; k=\frac{A}{2}(\xi-\frac{1}{\xi})
,\;\; \xi+\overline{\xi} \neq 0
\nonumber
\end{eqnarray}
Then $\xi=Re^{-i\alpha}$. Now:
\begin{equation}
\phi=\frac{1}{2}(\ae x-\gamma t)+i\frac{1}{2}(kx -\omega t)
\nonumber
\end{equation}
Here
\begin{eqnarray}
\ae=A(R-\frac{1}{R})\cos(\alpha) ,\;\; \gamma=-\frac{A^2}{2}(R^2+\frac{1}{R^2})\sin(2\alpha) \nonumber\\
k=A(R+\frac{1}{R})\sin(\alpha) ,\;\; \omega=\frac{A^2}{2}(R^2-\frac{1}{R^2})\cos(2\alpha) \nonumber
\end{eqnarray}
After the proper choice of $C$ we finish with
\begin{eqnarray}
\varphi=
\frac{Ae^{2i\alpha}}{2}\biggl(
\frac
{
2\cos(2\alpha)\cosh(u+w)
-\frac{1}{a}(R^2+\frac{1}{R^2})\cos(v)
}
{
\cosh(u+w)-\frac{1}{a}\cos(v)
}
+
\nonumber\\
i\frac
{
2\sin(2\alpha)\sinh(u+w)
+(R^2-\frac{1}{R^2})\sin(v)
}
{
\cosh(u+w)-\frac{1}{a}\cos(v)
}
\biggr)
\nonumber
\end{eqnarray}
Here
\begin{eqnarray}
u=\phi+\phi^*=\ae x-\gamma t ,\;\; v=\phi-\phi^*=i(k x-\omega t) \nonumber\\
a=\frac{1+R^2}{2R\cos(\alpha)} ,\;\; w=\ln(a) \nonumber\\
\end{eqnarray}

Note that
\begin{eqnarray}
\varphi \ra A \;\;\;\;\;\text{at}\;\;\;x \ra -\infty \nonumber\\
\varphi \ra Ae^{4i\alpha}\;\;\;\;\;\text{at}\;\;\;x \ra +\infty \nonumber\\
|\varphi|^{2}=A^{2} \;\;\;\;\;\text{at}\;\;\;x \ra \pm\infty\nonumber
\end{eqnarray}

\begin{figure}[htb]
\centering
\includegraphics[width=3.2in]{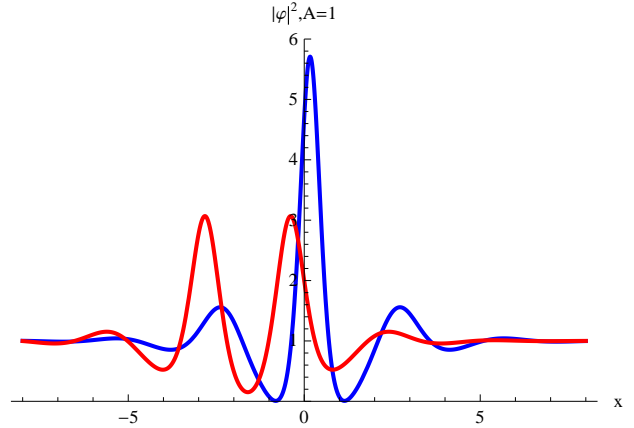}
\caption{\label{Solution1}
Typical solitonic solution at the moment of maximum (blue) and minimum (red) of amplitude. ($R=2,\alpha=\frac{5}{16}\pi$)
}
\end{figure}

Fig.~\ref{Solution1} show a typical solitonic solution at the moment of maximum and minimum amplitude.

\begin{figure}[htb]
\centering
\includegraphics[width=3.2in]{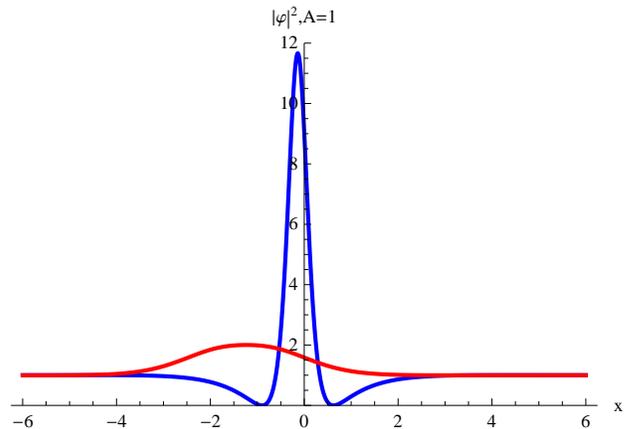}
\caption{\label{Kuznetsov-like}
Kuznetsov-like solution at the moment of maximum (blue) and minimum (red) of amplitude ($R=2,\alpha=\frac{\pi}{12}$).
}
\end{figure}

In our notation, we obtain Kuztensov's solution in the case of real $\xi$. When the pole is near the real axis, our solution looks like a Kuznetsov soliton moving with constant speed. This speed tends to zero in the limit $Im(\xi) \ra 0$. Thus, there is a broad area of Kuznetsov-like solutions near the real $\xi$ axis. Fig.~\ref{Kuznetsov-like} shows a typical solution corresponding to this case. The blue curve corresponds to the moment of maximum value of $|\varphi|^2$, while the red one corresponds to its minimal value.

\begin{figure}[htb]
\centering
\includegraphics[width=3.2in]{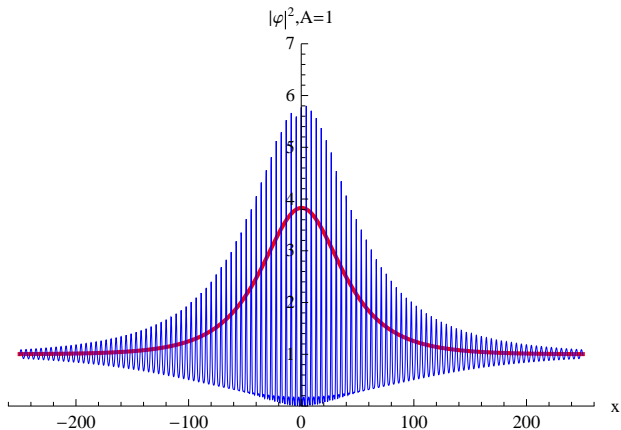}
\caption{\label{Soliton and Envelope}
Soliton and its envelope. ($R=1.01,\alpha=\frac{\pi}{4}$)
}
\end{figure}

The Akhmediev case appears for $|\xi|^{2}=1$. Near this area the solution is a wave train moving with constant speed and without changing its shape. We obtain the approximate expression for the envelope $s(x,t)$ of this wave train in the case of $Arg(\xi)=\frac{\pi}{4}$:
\begin{equation}
|s|^{2}=A\biggl(1+
Q\frac
{
\cosh(u+w)
}
{
(a^{2}\cosh^{2}(u+w)-4)^{\frac{3}{2}}
}\biggr) \label{Envelope}
\end{equation}
Here
\begin{eqnarray}
Q=a(6+R^{4}+\frac{1}{R^{4}}-4a^{2}) \nonumber\\
\end{eqnarray}
For an arbitrary value of $Arg(\xi)$ the envelope $s$ behaves similarly but the expression for it is more complicated. Fig.~\ref{Soliton and Envelope} shows a typical solitonic solution near $|\xi|^{2}=1$ and its envelope as calculated by (\ref{Envelope}).

The speed and size of the wave train increases without bound in the limit case when $\xi$ tends to the circle $|\xi|^{2}=1$. Thus the Akhmediev solution can be understood as a special case when the size and speed of the wave train tends to infinity.

Using the presented method one can construct multisoliton solutions with different values $\xi_{1},..,\xi_{n}$ of the uniformizing parameter $\xi$. By taking all $\xi_{i} \ra 1$ one can obtain "multi-instanton" solutions found recently in articles ~\cite{Matveev2010,Matveev2011}.

{\it --Conclusion --} We have discovered new solitonic solutions of the NLSE. This discovery essentially increases the number of candidate solutions for the analytic description of rogue waves. We expect that similar solutions exist in more exact models than NLSE. Since the found solutions change the condensate phase they can hardly be used for this purpose directly. However, the two-soliton solutions with complex conjugated values of poles ($\eta,\overline{\eta}$) are relevant candidates which could compete with homoclinic instanton-like solutions.

\begin{acknowledgments}
The authors express deep gratitude to Dr. E.A. Kuznetsov for helpful discussions.
This work was supported by:
The Grant of the Government of Russian Federation designed to support scientific projects implemented under the supervision of leading scientists at Russian institutions of higher education (No. 11.G34.31.0035), ONR Grant No. N00014-10-1-0991, NSF Grant DMS 0404577, and by the Grant "Leading Scientific Schools of Russia" (No. 6885.2010.2).
\end{acknowledgments}

\end{document}